\documentclass[%
print,
 amsmath,amssymb,
 aps,
]{revtex4}

\usepackage[sc]{mathpazo}
\usepackage{graphicx}
\usepackage{dcolumn}
\usepackage{bm}
\usepackage{dsfont}
\usepackage[landscape,papersize={297.1mm,210mm},left=1.9cm,right=1.6cm,top=2.7cm,bottom=2.8cm]{geometry}
\usepackage[                   
            pdfstartview=FitH,
            colorlinks, 
            pdfborder=001,   
            linkcolor=blue,
            anchorcolor=blue,
            citecolor=blue,
            urlcolor=blue
            ]{hyperref}

\begin{document}
\title{Relations among $k$-ME concurrence, negativity, polynomial invariants, and tangle}

\author{Limei~Zhang,~Ting~Gao}
\email{gaoting@hebtu.edu.cn}
\affiliation {College of Mathematics and Information Science,\\
 Hebei Normal University, Shijiazhuang 050024, China}
\author{Fengli~Yan}
\email{flyan@hebtu.edu.cn}
\affiliation {College of Physics Science and Information Engineering, Hebei Normal University, Shijiazhuang 050024, China}

\begin{abstract}
The $k$-ME concurrence as a measure of multipartite entanglement (ME) unambiguously detects all $k$-nonseparable states in arbitrary dimensions, and satisfies many important properties of an entanglement measure. Negativity is a simple computable bipartite entanglement measure.
Invariant and tangle are useful tools to study the properties of the quantum states. In this paper we mainly investigate  the internal relations among the $k$-ME concurrence, negativity, polynomial invariants, and tangle. Strong links between $k$-ME concurrence and negativity as well as between $k$-ME concurrence and polynomial invariants are derived. We obtain the quantitative relation between $k$-ME ($k$=$n$) concurrence and negativity for all $n$-qubit states, give a  exact value of the $n$-ME concurrence for the mixture of $n$-qubit GHZ states and white noise, and derive an  connection between $k$-ME concurrence and tangle for $n$-qubit W state. Moreover, we find that for any $3$-qubit pure state  the $k$-ME concurrence ($k$=2, 3) is  related to  negativity, tangle and polynomial invariants, while for $4$-qubit states the relations between  $k$-ME concurrence (for $k$=2, 4) and negativity, and between  $k$-ME concurrence and polynomial invariants also exist.  Our work provides clear quantitative connections between $k$-ME concurrence and negativity, and between $k$-ME concurrence and polynomial invariants.
\end{abstract}

\pacs{ 03.67.Mn, 03.65.Ud, 03.67.-a}

\maketitle

\section{Introduction}
Quantum entanglement \cite{RMP81.865} underlies most nonclassical  features of quantum physics, meanwhile  it is also a key resource for
showing quantum superiority in various information processing tasks, such as quantum cryptography \cite{PRL67.661}, quantum teleportation \cite{PRL70.1895,EPL84.50001}, quantum dense coding \cite{PRL69.2881}, and quantum metrology \cite{Nature2010}. Over the past three decades,  quantum entanglement has attracted unprecedented attentions. Great progress has been made in the characterization and quantification of entanglement in quantum-information science, and a number of entanglement measures are proposed for bipartite states \cite{ JPA.47.424005}.\\

Concurrence is one of the most well-accepted entanglement measures. Bennett \textit{et al.} \cite{7} first introduced it as an auxiliary tool to calculate the entanglement of formation for Bell-diagonal two-qubit states. Subsequently, concurrence was proposed  as an entanglement measure for two-qubit states by Wootters \textit{et al.} who derived computable formulas for concurrence and entanglement of formation in the two-qubit case \cite{8,9}. Later, generalization to the bipartite higher-dimensional systems \cite{10} as well as for multipartite systems \cite{11} were presented.
Gao \textit{et al.} introduced a generalized concurrence, $k$-ME concurrence \cite{18}, as a  multipartite entanglement (ME) measure for arbitrary high dimensional multipartite systems. The $k$-ME concurrence is well-defined, and satisfies important characteristics of an entanglement measure, such as entanglement monotone, vanishing on all $k$-separable states, invariant under local unitary transformations, subadditivity, convexity, and strictly greater than zero for  $k$-nonseparable states, \textit{etc} \cite{18}.  $k$-ME concurrence as a measure of multipartite $k$ inseparability, can unequivocally detects all $k$-nonseparable states in arbitrary dimensions \cite{17,40}.  Subsequently, Gao \textit{et al.} gave a lower bound of $k$-ME concurrence by using the permutation invariance part of a state, which can be applied to arbitrary multipartite systems \cite{41}.\\

 Negativity is another celebrated  bipartite entanglement measure for pure and mixed states  because of its simplicity and versatility. It was firstly shown by Vidal and Werner in 2002 \cite{PRA65.032314}. For a general  state  $\rho$   with  positive-semidefinite unit-trace  on  $\mathcal{H}_{A} \bigotimes \mathcal{H}_{B}$, the negativity is defined as: $N(\rho)=\frac{1}{2}(||\rho^{T_A}||_1-1)$, where $\rho^{T_A}$ denotes the partial transpose of $\rho$ with respect to the subsystem $A$, and $||X||_1=\mathrm{Tr}\sqrt{X^{\dag}X}$ is the trace norm (\textit{i.e.} the sum of all singular values) of matrix $X$. Note that the negativity of $\rho$ is also the absolute value of the sum of the negative eigenvalues of  partial transpose $\rho^{T_A}$ \cite{PRA65.032314}.
For the $ 2 \bigotimes 2$ and $2 \bigotimes 3$ systems, it vanishes if and only if $\rho_{AB}$ is separable \cite{PRA61.062313}.
   Nevertheless, the negativity fails to recognize entanglement in a positive partial transpose (PPT) state \cite{PRL80.5239}. To compensate for this insufficiency, Lee \textit{et al.} presented an extended negativity on mixed states, which is called the convex-roof extended negativity (CREN) \cite{PRA68.062304}. The CREN is nonzero iff the state is entangled \cite{PRA68.062304}. Negativity is an entanglement monotone \cite{JM047}, and always a lower bound to the CREN. Consider an $n$-party state $\rho$ on Hilbert space $\mathcal{H} = \mathcal{H}_{A_1} \bigotimes \mathcal{H}_{A_2} \bigotimes\cdots \bigotimes \mathcal{H}_{A_n}$, one may use the global negativity \cite{PRA77.042117,JPA.47.424005} to  measure the entanglement of subsystem ${A_i}$ with its complement in a bipartite split of the composite system.\\

Classification of entanglement is a significant task of quantum information, which can help in recognizing similarity between different entangled states \cite{pra62062314}. Meanwhile it  is useful to boost the practicabilities of quantum information protocols.  Entanglement classification  arouses  continuous attentions \cite{PRA.87,PRA.88,JPA.50.2017,pra82032121,prl108050501,PRA.58.1833}.  Polynomial invariants are important factors in research on entanglement classification and quantification. Linden \textit{et al.}  had shown that for a state there is an infinite set of polynomial functions, which is invariant under local unitary (LU) transformations \cite{20}. Then polynomial invariants are regarded as indicators on the space of entanglement types. It is clear that a measure of entanglement is a function on the space of states of a multiparticle system which is invariant under local unitary operators \cite{JMP43}. That is , any measure of entanglement for multiparticles must be a function of the invariants. It means that polynomial invariants are close connected to entanglement measures.\\

Tangle is an invariant but is not a valid entanglement measure. It is an essential index to describe the properties of a state. The 3-tangle is maximal for the GHZ state but vanishes for the W state. This means that the three-tangle is not suitable as a measure of entanglement. The 3-tangle reveals the existence of two inequivalent kinds of genuine tripartite entanglement for pure three-qubit states \cite{pra62062314}.\\

The $k$-ME concurrence, negativity, tangle, and polynomial invariant are all used to characterize quantum entanglement. It is then legitimate to ask what is the internal link between each other? In this paper we provide mathematically rigorous approach to resolve the above question.\\

 This paper is organized as follows. In Sec.II we review the definitions of $k$-ME concurrence, negativity, and tangle. In Sec.III we give a general relation between $k$-ME ($k$=$n$) concurrence and negativity for $n$-qubit states, and obtain a  exact value of the $n$-ME concurrence for the mixture of $n$-qubit GHZ states and white noise.
  Then we derive quantitative connections among $k$-ME concurrence, negativity, tangle, and polynomial invariant for 3-qubit states in Sec.IV, and  for the nine families of four-qubit states under stochastic local operations and classical communication (SLOCC) in Sec.V.  In Sec.VI, we illustrate the link between $k$-ME concurrence and tangle for $n$-qubit W state. Finally, we summarize our results in Sec.VII.

\section{Preliminaries}
Before stating the main results, we firstly introduce some necessary concepts and notations that will be used in this paper. Let $\mathcal{H} = \mathcal{H}_1 \bigotimes \mathcal{H}_2 \bigotimes\cdots \bigotimes \mathcal{H}_n$ be the state space, where $\mathcal{H}_i$  is a complex Hilbert space
with $dim\mathcal{H}_i= d_i~(i=1,2,\cdots,n)$.
 A $k$-partition $A_1 | A_2 | \cdots | A_k$~of $n$ quantum subsystems means that the set $\{A_1,A_2,\cdots,A_k\}$ is a collection of pairwise disjoint nonempty subsets of  $\{1, 2, \cdots, n\}$, and the union of all sets in $\{A_1, A_2, \cdots, A_k\}$ is $\{1,2,\cdots,n\}$ (disjoint union~$\cup_{i=1}^{k}A_{i}=\{1,2,\cdots,n\}$).
An $n$-partite pure state~$|\psi\rangle \in \mathcal{H}_1 \bigotimes \mathcal{H}_2 \bigotimes\cdots \bigotimes \mathcal{H}_n$~is said to be $k$-separable \cite{18,40}, if there exists a $k$-partition~$A_1|A_2|\cdots|A_k$ of $\{1, 2, \cdots, n\}$ such that
\begin{equation*}
|\psi\rangle=|\psi_1\rangle_{A_1}|\psi_2\rangle_{A_2}\cdots|\psi_k\rangle_{A_k},
\end{equation*}
where $|\psi_i\rangle_{A_i}$ is a pure state of $\mathcal{H}_{A_i}$.
An $n$-partite mixed state $\rho$ is $k$-separable if it can be written as a convex combination of $k$-separable pure states,\begin{equation*}\rho=\sum\limits_{i}p_i|\psi_i\rangle \langle\psi_i|,\end{equation*}
where $p_i \geq 0,~ \sum_ip_i=1$. Here $|\psi_i\rangle$ may be $k$-separable with respect to different $k$-partitions.\\

$k$-ME concurrence \cite{18} is an entanglement measure of arbitrary high-dimensional multipartite systems.
For an $n$-partite pure state~$|\psi\rangle \in \mathcal{H}_1\otimes\mathcal{H}_2\otimes\cdots\otimes\mathcal{H}_n$, the $k$-ME concurrence of $|\psi\rangle$ is defined as
\begin{equation}\label{0}
C_{k-\mathrm{ME}}(|\psi\rangle)=\min\limits_{A}\sqrt{2\left(1-\frac{\sum\limits_{t=1}^k\textrm{Tr}(\rho^2_{A_t})}{k}\right)}
=\min\limits_{A}\sqrt{\frac{2\sum\limits_{t=1}^k\left[1-\textrm{Tr}(\rho^2_{A_t})\right]}{k}},
\end{equation}
 where~the minimum is taken over all possible $k$-partitions $A=A_1|A_2|\cdots|A_k$ of $\{1,2,\cdots,n\}$ and
 $\rho_{A_t}=\mathrm{Tr}_{\bar{A_t}}(|\psi\rangle\langle\psi|)$~denotes the reduced density matrix of
subsystem $A_t$ ($\bar{A_t}$ is the complement of $A_t$ in $\{1,2,\cdots,n\}$).\\

Suppose  that $\rho$ is an $n$-partite state on $\mathcal{H}_1\otimes\mathcal{H}_2\otimes\cdots\otimes\mathcal{H}_n$ with $dim \mathcal{H}_p=d_p$.   Global negativity \cite{PRA77.042117,JPA.47.424005} is
\begin{equation}
N^{p}=\frac{1}{d_{p}-1}(||\rho^{T_{p}}||_{1}-1)=-\frac{2}{d_{p}-1}\sum\limits_{i}\lambda_{i}^{p_{-}},\end{equation}
herein $\rho^{T_{p}}$ is the partial transpose with respect to the $p$th subsystem, $||\rho||_{1}$ is the trace norm of $\rho$,  and $\lambda_{i}^{p_-}$ are the negative eigenvalues of~$\rho^{T_{p}}$.  Note that we only consider the situation $d_{p}=2$ ($p=1,2,\cdots,n$) throughout the paper, and thereby the formula (2) can be rewritten
\begin{equation}\label{60}
N^{p}=||\rho^{T_{p}}||_{1}-1=-2\sum_{i}\lambda_{i}^{p_{-}}
\end{equation}
for qubit systems.\\

For a two-qubit pure state $|\psi_{AB}\rangle$, the tangle (or one tangle) is defined as \cite{37,pra92042307}
\begin{equation}
\tau(|\psi\rangle_{A|B})=4 \det \rho_A,
\end{equation}
where $\rho_A= \mathrm{Tr}_{B}|\psi\rangle_{AB}\langle\psi|$ is the reduced density matrix. For a two qubit mixed state $\rho_{AB}$, with pure state decomposition $\rho_{AB}=\sum_i p_i|\psi_i\rangle\langle\psi_i|$, the tangle (or two tangle) is
\begin{equation}
\tau(\rho_{A|B})=[\min_{\{p_i,|\psi_i\rangle\}}\sum_ip_i\sqrt{\tau(|\psi_i\rangle_{A|B})}]^2,
\end{equation}
where the minimization is taken over all possible  pure state decompositions.\\

For any three-qubit pure state $|\psi_{ABC}\rangle$, three-tangle \cite{37} is defined as
\begin{equation}
\tau(|\psi\rangle_{A|B|C})=\tau(|\psi\rangle_{A|BC})-\tau(\rho_{A|B})-\tau(\rho_{A|C}),
\end{equation}
where $\tau(|\psi\rangle_{A|BC})$ is the one-tangle of pure state $|\psi_{ABC}\rangle$ quantifying the bipartite entanglement between subsystems $A$ and $BC$.   $\tau(\rho_{A|B})$ (or $\tau(\rho_{A|C})$) is the  two-tangle of  reduced states $\rho_{AB}= \mathrm{Tr}_C (|\psi\rangle_{ABC}\langle\psi|)$ (or $\rho_{AC}= \mathrm{Tr}_B (|\psi\rangle_{ABC}\langle\psi|$)). The three-tangle is a good measure of genuine three-qubit entanglement.

\par
\section{ relations between the $k$-ME concurrence and negativity for $n$-qubit states}
A pure $d_A\bigotimes d_B$ $(d_A\leq d_B)$ quantum state $|\psi\rangle$ has the standard Schmidt decomposition \cite{12}
\begin{equation*}
|\psi\rangle=\sum_{i}\sqrt{\lambda_i}|i_Ai_B\rangle,
\end{equation*}
where $\sqrt{\lambda_i}$ $(i=1,2,\cdots,d_A)$ are the Schmidt coefficients, $|i_A\rangle$ and $|i_B\rangle$ are the orthonormal basis in $\mathcal{H}_A$ and $\mathcal{H}_B$ respectively. $\rho=|\psi\rangle\langle\psi|=\sum_{i j}\sqrt{\lambda_i\lambda_j}|i_Ai_B\rangle\langle j_Aj_B|$ is the density operator of $|\psi\rangle$. The concurrence of $|\psi\rangle$ is
\begin{equation*}
C(|\psi\rangle)=\sqrt{2(1-\mathrm{Tr}\rho_A^2)}=\sqrt{2(1-\sum_i\lambda_i^2)}=\sqrt{4\sum_{i<j} \lambda_i\lambda_j}.
\end{equation*}
The negativity of $|\psi\rangle$ is
\begin{equation*}
N(|\psi\rangle)=||\rho^{T_{A}}||_{1}-1=2\sum_{i<j}\sqrt{\lambda_i\lambda_j}.
\end{equation*}
If the Schmidt rank (the number of non-zero values $\lambda_i$) of $|\psi\rangle$ is 2, we obtain a equality relation
\begin{equation}
N(|\psi\rangle)=C(|\psi\rangle).
\end{equation}

For any $n$-qubit pure state $\rho_{1 2 \cdots n}$, $C_{p|A}(\rho)$ is the concurrence of the state $\rho_{1 2 \cdots n}$ which is viewed as a bipartite state under the partition $p$ and $A$. Here $A=\{1,2,\cdots,n\}\backslash\{p\}$ and $p=1,2,\cdots,n$.
Then the global negativity and concurrence have relation
\begin{equation}\label{2}
N^{p}(\rho)=C_{p|A}(\rho)=\sqrt{2[1-\mathrm{Tr}(\rho_p^2)]}.
\end{equation}
Thus, we obtain the relation between $k$-ME concurrence ($k$=$n$) and negativity for all the $n$-qubit pure states
 \begin{equation}
C_{n-\mathrm{ME}}(|\psi\rangle)=\sqrt{\frac{(N^{1})^2+(N^{2})^2+\cdots+(N^{n})^2}{n}}.
\end{equation}
While for any $n$-qubit mixed states $\rho$, we have
 \begin{equation}\label{n-ME-LowerBound}
C_{n-\mathrm{ME}}(\rho)\geq\sqrt{\frac{(N^{1})^2+(N^{2})^2+\cdots+(N^{n})^2}{n}}.
\end{equation}

This can be seen as follows: Let
$\rho=\sum_ip_i|\psi_i\rangle\langle\psi_i|$
be an optimal decomposition such that
\begin{equation*}
C_{n-\mathrm{ME}}(\rho)=\sum_ip_iC_{n-\mathrm{ME}}(|\psi_i\rangle).
\end{equation*}
Then we get
\begin{equation}
\begin{array}{rl}
C_{n-\mathrm{ME}}(\rho)=&\frac{1}{\sqrt{n}}\sum_ip_i\{\sum_{j=1}^{n}[N^{j}(|\psi_i\rangle)]^2\}^{\frac{1}{2}}\\\\
\geq&\frac{1}{\sqrt{n}}\{\sum_{j=1}^{n}[\sum_ip_iN^j(|\psi_i\rangle)]^2\}^{\frac{1}{2}}\\\\
\geq&\frac{1}{\sqrt{n}}\{\sum_{j=1}^{n}[N^j(\rho)]^2\}^{\frac{1}{2}},
\end{array}
\end{equation}
which is the desired result. Here in the first inequality we have used the fact that, for nonnegative real numbers $x_{ij}$, $\sum_i(\sum_jx_{ij}^2)^{\frac{1}{2}}\geq[\sum_j(\sum_ix_{ij})^2]^{\frac{1}{2}}$, and the second inequality because of the convex property of negativity.\\

The lower bound of $n$-ME concurrence $C_{n-\text{ME}}(\rho)$ in Eq.(\ref{n-ME-LowerBound}) is tight.  The inequality (\ref{n-ME-LowerBound}) can be saturated not only for all pure $n$-qubit states but also for some mixed states.

\emph{Example.}  Consider the family of $n$-qubit states,  the $n$-qubit GHZ states mixed with white noise,
\begin{equation*}
\rho^{G}(t)=\frac{1-t}{2^{n}}\mathcal{I}+t|GHZ_n\rangle\langle GHZ_n|
\end{equation*}
where $|GHZ_n\rangle=\frac{1}{\sqrt{2}}(|00\cdots0\rangle+|11\cdots1\rangle)$.

If  $t\geq\frac{1}{2^{n-1}+1}$, then negativities of $\rho$ are
  \begin{equation}
N^{1}=N^{2}=\cdots=N^{n}=\frac{(2^{n-1}+1)t-1}{2^{n-1}}.
\end{equation}
Combining it with Eq.(\ref{n-ME-LowerBound}), there is
 \begin{equation}
C_{n-\mathrm{ME}}(\rho^{G}(t))\geq\frac{(2^{n-1}+1)t-1}{2^{n-1}},  ~~ t\in[\frac{1}{2^{n-1}+1},1],
\end{equation}
which shows that $\rho^{G}(t)$ is entangled (not $n$-separable, or not fully separable) if $t > \frac{1}{2^{n-1}+1}$  \cite{18}. Note that for $t=1$,  this lower bound is equal to the $n$-ME concurrence of the pure $n$-qubit GHZ state $|GHZ_n\rangle$.
This means that the criterion from Eq.(\ref{n-ME-LowerBound}) constitutes a
necessary and sufficient criterion for entanglement for the family of states $\rho^{G}(t)$, since it is known that  $\rho^{G}(t)$ are fully separable ($n$-separable) iff  $0\leq t\leq\frac{1}{2^{n-1}+1}$  \cite{PRA61042314,GaoEPJD2011}.
Because $n$-ME concurrence is a convex function \cite{18} and this bound coincides with the exact value on the points $t = \frac{1}{2^{n-1}+1}$ and $t = 1$,
the bound equals the exact value on the whole interval $t\in[\frac{1}{2^{n-1}+1},1]$, that is, we derive
\begin{equation}
C_{n-\mathrm{ME}}(\rho^{G}(t))=\frac{(2^{n-1}+1)t-1}{2^{n-1}},~~~~t\in[\frac{1}{2^{n-1}+1},1],
\end{equation}
for the mixture of $n$-qubit GHZ states and  white noise.

\section{Relations among the $k$-ME concurrence, negativity, tangle, and polynomial invariants for three qubit states}
In this section, we consider the system of three spin-$1/2$ particles, and study the relations among $k$-ME concurrence, negativity, tangle, and polynomial invariants  for 3-qubit states.
\subsection{Relationship between $k$-ME concurrence and negativity for 3-qubit states}
For any three-qubit pure state $|\psi\rangle \in \mathcal{C}^2_A \bigotimes \mathcal{C}^2_B \bigotimes \mathcal{C}^2_C $, from Eq.$\eqref{2}$ we obtain the global negativity of $|\psi\rangle$,
\begin{equation*}\label{4}\small
N^{A}=\sqrt{2[1-\mathrm{Tr}(\rho_A^2)]},~~~~N^{B}=\sqrt{2[1-\mathrm{Tr}(\rho_B^2)]},~~~~N^{C}=\sqrt{2[1-\mathrm{Tr}(\rho_C^2)]},
\end{equation*}
where $\rho_X$ ($X=A,B,C$) are the reduced density operators of $\rho=|\psi\rangle\langle\psi|$ with respect to the subsystem $X$.
By the definition of $k$-ME concurrence, one can easily get the following relations between $k$-ME concurrence ($k$=2, 3) and negativities for 3-qubit state $|\psi\rangle$
\begin{equation}
\begin{array}{rl}
C_{2-\mathrm{ME}}(|\psi\rangle) =&\min\{N^{A},N^{B},N^{C}\},\\\\
C_{3-\mathrm{ME}}(|\psi\rangle)=&\sqrt{\frac{(N^{A})^2+(N^{B})^2+(N^{C})^2}{3}}.\\
\end{array}
\end{equation}
\subsection{Relationship among $k$-ME concurrence, tangle, and polynomial invariants for 3-qubit states}
 In the invariant theory, for a 3-qubit state $|\psi\rangle$, there are one independent invariant of degree two \cite{36,jpa346787}
 \begin{equation*}
I_2=\langle\psi|\psi\rangle,
\end{equation*}
and four algebraically independent invariants of degree four
\begin{equation}
\begin{array}{rl}\label{8}
I^{(1)}_4=&\langle\psi|\psi\rangle^2,~~~I^{(2)}_4=~~\mathrm{Tr}(\rho_A^2),~~~ I^{(3)}_4=~~\mathrm{Tr}(\rho_B^2),~~~I^{(4)}_4=~~\mathrm{Tr}(\rho_C^2).
\end{array}
\end{equation}

  Sudbery \cite{36}~had shown that  these polynomial invariants have relations with all 2-tangles and the 3-tangle \cite{37,pra92042307}, such as
\begin{equation}
\begin{array}{c}\label{6}
  \tau_{AB}=1-I^{(2)}_4-I^{(3)}_4+I^{(4)}_4-\frac{1}{2} \tau_{ABC}, \\\\
  \tau_{AC}=1-I^{(2)}_4+I^{(3)}_4-I^{(4)}_4-\frac{1}{2} \tau_{ABC}, \\\\
   \tau_{BC}=1+I^{(2)}_4-I^{(3)}_4-I^{(4)}_4-\frac{1}{2} \tau_{ABC}.
\end{array}
\end{equation}
The combination of Eq.$\eqref{0}$, Eq.$\eqref{8}$,  and Eq.$\eqref{6}$ gives the formulae for $2$-ME concurrence and $3$-ME concurrence of three-qubit states in terms of all three 2-tangles and polynomial invariants
\begin{equation*}
\begin{array}{rl}
C_{2-\mathrm{ME}}(|\psi\rangle)=&\sqrt{2(I_2-\max\{I^{(m)}_4|_{m=2,3,4}\})},\\\\
C_{3-\mathrm{ME}}(|\psi\rangle)=&\sqrt{\frac{2}{3}(3I_2-I^{(2)}_4-I^{(3)}_4-I^{(4)}_4)}\\
=&\sqrt{\frac{2}{3}(\tau_{AB}+\tau_{AC}+ \tau_{BC})+\tau_{ABC}}.\\
\end{array}
\end{equation*}
\section{Relations among the $k$-ME Concurrence, negativity, and Polynomial Invariants for four-qubit states}

\subsection{Relationship between $k$-ME concurrence and polynomial invariants for 4-qubit states}
 Let $|\psi\rangle \in \mathcal{C}^2_A \bigotimes \mathcal{C}^2_B \bigotimes \mathcal{C}^2_C \bigotimes \mathcal{C}^2_D$ be a $4$-qubit~pure state, then there exists  one local invariant of degree two \cite{20}
\begin{equation}
I_2=\langle\psi|\psi\rangle,
\end{equation}
and seven degree-4 independent invariants
\begin{equation}
\begin{array}{rl}\label{7}
 I^{(1)}_4=&\mathrm{Tr}(\rho_D^2),~~~~~~~~I^{(2)}_4=\mathrm{Tr}(\rho_C^2),~~~~~~~~I^{(3)}_4=\mathrm{Tr}(\rho_B^2),\\\\
I^{(4)}_4=&\mathrm{Tr}(\rho_A^2),~~~~~~~~I^{(5)}_4=\mathrm{Tr}(\rho_{AD}^2),~~~~~I^{(6)}_4=\mathrm{Tr}(\rho_{BD}^2),\\\\
 I^{(7)}_4=&\mathrm{Tr}(\rho_{CD}^2).
\end{array}
\end{equation}
From Eqs.$\eqref{0}$ and $\eqref{7}$, we get following relations for a 4-qubit state
\begin{equation*}
\begin{array}{rl}
C_{2-\mathrm{ME}}(|\psi\rangle)=&\sqrt{2(I_2-\max\{I^{(m)}_4|_{m=1,2,\cdots,7}\})},\\\\
C_{4-\mathrm{ME}}(|\psi\rangle)=&\sqrt{\frac{4I_2-I^{(1)}_4-I^{(2)}_4-I^{(3)}_4-I^{(4)}_4}{2}},
\end{array}
\end{equation*}
and
\begin{equation*}\small
\begin{array}{rl}
C_{3-\mathrm{ME}}(|\psi\rangle)=&\min\{\sqrt{\frac{2}{3}(3I_2-I^{(4)}_4-I^{(3)}_4-I^{(7)}_4)},~\sqrt{\frac{2}{3}(3I_2-I^{(4)}_4-I^{(2)}_4-I^{(6)}_4)},\\
&~~~~~~~~~\sqrt{\frac{2}{3}(3I_2-I^{(4)}_4-I^{(1)}_4-I^{(5)}_4)},~\sqrt{\frac{2}{3}(3I_2-I^{(3)}_4-I^{(2)}_4-I^{(5)}_4)},\\
&~~~~~~~~\sqrt{\frac{2}{3}(3I_2-I^{(3)}_4-I^{(1)}_4-I^{(6)}_4)},~\sqrt{\frac{2}{3}(3I_2-I^{(2)}_4-I^{(1)}_4-I^{(7)}_4)}~\}.
\end{array}
\end{equation*}

\subsection{Relationship between $k$-ME concurrence and negativity for 4-qubit states}
For a normalized $4$-qubit state $\rho$, the $4$-ME concurrence and negativity have the relation
\begin{equation}
C_{4-\mathrm{ME}}(\rho)\geq\sqrt{\frac{(N^{1})^2+(N^{2})^2+(N^{3})^2+(N^{4})^2}{4}},
\end{equation}
with equality for all 4-qubit pure states.\\

Verstraete \cite{PRA65.2002.052112} investigated  the behavior of a single copy of a pure four-qubit state under the action
of stochastic local quantum operations assisted by classical communication (SLOCC), and proved  that there exist nine families of
states corresponding to nine different ways of entangling  for four-qubits:
\begin{small}
\begin{equation}
\begin{array}{l}
  G_{abcd}=\frac{a+d}{2}(|0000\rangle+|1111\rangle)+\frac{a-d}{2}(|0011\rangle+|1100\rangle)
  +\frac{b+c}{2}(|0101\rangle+|1010\rangle)+\frac{b-c}{2}(|0110\rangle+|1001\rangle),\\\\
L_{abc_{2}}=\frac{a+b}{2}(|0000\rangle+|1111\rangle)+\frac{a-b}{2}(|0011\rangle+|1100\rangle)+c(|0101\rangle+|1010\rangle)+|0110\rangle,\\\\
L_{a_{2}b_{2}}=a(|0000\rangle+|1111\rangle)+b(|0101\rangle+|1010\rangle)+|0110\rangle+|0011\rangle,\\\\
L_{ab_{3}}=a(|0000\rangle+|1111\rangle)+\frac{a+b}{2}(|0101\rangle+|1010\rangle)+\frac{a-b}{2}(|0110\rangle+|1001\rangle)\\
~~~~~~~~~+\frac{\mathrm{i}}{\sqrt{2}}(|0001\rangle+|0010\rangle+|0111\rangle+|1011\rangle),\\\\
L_{a_{4}}=a(|0000\rangle+|0101\rangle+|1010\rangle+|1111\rangle)+(\mathrm{i}|0001\rangle+|0110\rangle-\mathrm{i}|1011\rangle),\\\\
L_{a_{2}0_{3\bigoplus\bar{1}}}=a(|0000\rangle+|1111\rangle)+|0011\rangle+|0101\rangle+|0110\rangle,\\\\
L_{0_{5\bigoplus\bar{3}}}=|0000\rangle+|0101\rangle+|1000\rangle+|1110\rangle,\\\\
L_{0_{7\bigoplus\bar{1}}}=|0000\rangle+|1011\rangle+|1101\rangle+|1110\rangle,\\\\
L_{0_{3\bigoplus\bar{1}}0_{3\bigoplus\bar{1}}}=|0000\rangle+|0111\rangle,
\end{array}
\end{equation}
\end{small}
where~$a,~b,~c,$ and $d$~are  the same as in \cite{PRA65.2002.052112}.\\

(\uppercase\expandafter{\romannumeral 1})~~For $L_{0_{3\bigoplus\bar{1}}0_{3\bigoplus\bar{1}}}$, let $$|\psi_{9}\rangle=\frac{L_{0_{3\bigoplus\bar{1}}0_{3\bigoplus\bar{1}}}}{||L_{0_{3\bigoplus\bar{1}}0_{3\bigoplus\bar{1}}}||}=\frac{1}{\sqrt{2}}(|0000\rangle+|0111\rangle).$$
Obviously, $|\psi_{9}\rangle$ is a $2$-separable state, thus we have
\begin{equation}\label{11}
\mathrm{C}_{2-\mathrm{ME}}(|\psi_{9}\rangle)=0.\end{equation}

Note that
\begin{equation}\small
\begin{array}{rl}\label{10}
\sqrt{\frac{2}{3}(1-\mathrm{Tr}\rho_1^2+1-\mathrm{Tr}\rho_2^2+1-\mathrm{Tr}\rho_{34}^2)}=&\sqrt{\frac{2}{3}},\\
\sqrt{\frac{2}{3}(1-\mathrm{Tr}\rho_1^2+1-\mathrm{Tr}\rho_3^2+1-\mathrm{Tr}\rho_{24}^2)}=&\sqrt{\frac{2}{3}},\\
\sqrt{\frac{2}{3}(1-\mathrm{Tr}\rho_1^2+1-\mathrm{Tr}\rho_4^2+1-\mathrm{Tr}\rho_{23}^2)}=&\sqrt{\frac{2}{3}},\\
\sqrt{\frac{2}{3}(1-\mathrm{Tr}\rho_2^2+1-\mathrm{Tr}\rho_3^2+1-\mathrm{Tr}\rho_{14}^2)}=&1,\\
\sqrt{\frac{2}{3}(1-\mathrm{Tr}\rho_2^2+1-\mathrm{Tr}\rho_4^2+1-\mathrm{Tr}\rho_{13}^2)}=&1,\\
\sqrt{\frac{2}{3}(1-\mathrm{Tr}\rho_3^2+1-\mathrm{Tr}\rho_4^2+1-\mathrm{Tr}\rho_{12}^2)}=&1,
\end{array}
\end{equation} where $\rho=|\psi_{9}\rangle\langle\psi_{9}|$ , and  $\rho_A$ is the reduced density operator of subsystem $A$$\subseteq\{1,2,3,4\}$, then $3$-ME concurrence of $|\psi_{9}\rangle$ is the minimum of $\label{eqrl}$ Eq.$\eqref{10}$,
that is \begin{equation*}
C_{3-\mathrm{ME}}(|\psi_{9}\rangle)=\sqrt{\frac{2}{3}}.\end{equation*}
According to definition Eq.$\eqref{0}$, 4-ME concurrence is
\begin{equation}
\begin{array}{rl}\label{12}
C_{4-\mathrm{ME}}(|\psi_{9}\rangle)=&\sqrt{\frac{2}{4}(1-\mathrm{Tr}\rho_1^2+1-\mathrm{Tr}\rho_2^2+1-\mathrm{Tr}\rho_3^2+1-\mathrm{Tr}\rho_4^2)}\\
=&\sqrt{\frac{2}{4}(0+\frac{1}{2}+\frac{1}{2}+\frac{1}{2})}=\frac{\sqrt{3}}{2}.
\end{array}
\end{equation}
The negativities of $|\psi_{9}\rangle$ are
\begin{equation}
\begin{array}{rl}\label{13}
N^{1}=&||\rho^{T_{1}}||_{1}-1=0,~~~~~~~~~~N^{2}=||\rho^{T_{2}}||_{1}-1=1,\\\\
N^{3}=&||\rho^{T_{3}}||_{1}-1=1,~~~~~~~~~~N^{4}=||\rho^{T_{4}}||_{1}-1=1.
\end{array}
\end{equation}
From Eqs.$\eqref{11}$ and $\eqref{13}$, one can get the following relation for $|\psi_{9}\rangle$
\begin{equation}
\begin{array}{rl}
C_{2-\mathrm{ME}}(|\psi_{9}\rangle)=\min\{N^{1},N^{2},N^{3},N^{4}\}.
\end{array}
\end{equation}

(\uppercase\expandafter{\romannumeral 2})~~We show the relation between $2$-ME concurrence and negativity for $\frac{L_{0_{7\bigoplus\bar{1}}}}{||L_{0_{7\bigoplus\bar{1}}}||}$ in an analogous way.
Suppose
\begin{equation*}
|\psi_{8}\rangle=\frac{L_{0_{7\bigoplus\bar{1}}}}{||L_{0_{7\bigoplus\bar{1}}}||}=\frac{1}{2}(|0000\rangle+|1011\rangle+|1101\rangle+|1110\rangle).
\end{equation*}
Since there are seven $2$-partitions for four qubits, then $2$-ME concurrence of $\rho=|\psi_{8}\rangle\langle\psi_{8}|$ is the minimum of $\sqrt{1-\mathrm{Tr}(\rho^2_{A_t})+1-\mathrm{Tr}(\rho^2_{\bar{A_t}})}$, taking over all possible $2$-partitions.  We can easily obtain
\begin{equation}\label{15}
C_{2-\mathrm{ME}}(|\psi_{8}\rangle)=\frac{\sqrt{3}}{2}.
\end{equation}
By Eq.$\eqref{0}$, one also has
\begin{equation}
\begin{array}{rl}\label{16}
C_{3-\mathrm{ME}}(|\psi_{8}\rangle)=&1,\\\\
C_{4-\mathrm{ME}}(|\psi_{8}\rangle)=&\frac{\sqrt{15}}{4}.
\end{array}
\end{equation}
From Eq.$\eqref{60}$, there are
\begin{equation}
\begin{array}{rl}\label{17}
N^{1}=&2\Sigma_{i}|\lambda^{1_{-}}_{i}|=\frac{\sqrt{3}}{2},~~~~~~~~~~~N^{2}=2\Sigma_{i}|\lambda^{2_{-}}_{i}|=1,\\\\
N^{3}=&2\Sigma_{i}|\lambda^{3_{-}}_{i}|=1,~~~~~~~~~~~~~N^{4}=2\Sigma_{i}|\lambda^{4_{-}}_{i}|=1,\\
\end{array}
\end{equation}where $\lambda^{p_{-}}_{i}$ is the negative eigenvalue of~$\rho^{T_{p}}$ ( $\rho^{T_{p}}$ is the partial transpose of $\rho=|\psi_{8}\rangle\langle\psi_{8}|$ with respect to the $p$th subsystem ).
Combining Eqs.$\eqref{15}$ and $\eqref{17}$,  there is
\begin{equation}
\begin{array}{rl}
C_{2-\mathrm{ME}}(|\psi_{8}\rangle)=\min\{N^{1},N^{2},N^{3},N^{4}\}.
\end{array}
\end{equation}

(\uppercase\expandafter{\romannumeral 3})~~For 4-qubit state $L_{0_{5\bigoplus\bar{3}}}$,  computing the $k$-ME concurrence and negativity, then a similar relation can be given.

 Let $$|\psi_{7}\rangle=\frac{L_{0_{5\bigoplus\bar{3}}}}{||L_{0_{5\bigoplus\bar{3}}}||}=\frac{1}{2}(|0000\rangle+|0101\rangle+|1000\rangle+|1110\rangle).$$

After some simple calculations, one have
\begin{equation}
\begin{array}{rl}
C_{2-\mathrm{ME}}(|\psi_{7}\rangle)=&\frac{\sqrt{3}}{2},\\\\
C_{3-\mathrm{ME}}(|\psi_{7}\rangle)=&\sqrt{\frac{5}{6}},\\\\
C_{4-\mathrm{ME}}(|\psi_{7}\rangle)=&\frac{\sqrt{13}}{4}.
\end{array}
\end{equation}
The negativities of $|\psi_{7}\rangle$ are
\begin{equation}
N^{1}=\frac{\sqrt{3}}{2},~~~N^{2}=1,~~~N^{3}=\frac{\sqrt{3}}{2},~~~N^{4}=\frac{\sqrt{3}}{2}.
\end{equation}
It follows that
\begin{equation}
\begin{array}{rl}
C_{2-\mathrm{ME}}(|\psi_{7}\rangle)=&\min\{N^{1},N^{2},N^{3},N^{4}\}.
\end{array}
\end{equation}

(\uppercase\expandafter{\romannumeral 4})~~We compute the $k$-ME concurrence and negativity of $L_{a_20_{3\bigoplus\bar{1}}}$. Let
\begin{equation*}
|\psi_{6}\rangle=\frac{L_{a_20_{3\bigoplus\bar{1}}}}{||L_{a_20_{3\bigoplus\bar{1}}}||}=\frac{1}{\sqrt{2|a|^2+3}}(a|0000\rangle+a|1111\rangle+|0011\rangle+|0101\rangle+|0110\rangle).
\end{equation*}

Through some tedious computing, one gets
\begin{equation}
\begin{array}{rl}\label{30}
C_{2-\mathrm{ME}}(|\psi_{6}\rangle)=&\sqrt{1-\frac{9}{(2|a|^2+3)^2}},\\\\
C_{3-\mathrm{ME}}(|\psi_{6}\rangle)=&\sqrt{1-\frac{11-4|a|^2}{3(2|a|^2+3)^2}},\\\\
C_{4-\mathrm{ME}}(|\psi_{6}\rangle)=&\sqrt{1-\frac{3}{(2|a|^2+3)^2}}.
\end{array}\end{equation}
The negativities of $|\psi_{6}\rangle$ are
\begin{equation}
\label{31}\small
N^{1}=\sqrt{1-\frac{9}{(2|a|^2+3)^2}},~~~N^{2}=N^{3}=N^{4}=\sqrt{1-\frac{1}{(2|a|^2+3)^2}}.
\end{equation}
From Eqs.$\eqref{30}$ and $\eqref{31}$, we obtain
\begin{equation}
\begin{array}{rl}\small
C_{2-\mathrm{ME}}(|\psi_{6}\rangle)=\min\{N^{1},N^{2},N^{3},N^{4}\}.
\end{array}
\end{equation}

(\uppercase\expandafter{\romannumeral 5})~~For $L_{a_4}$, suppose
$$|\psi_{5}\rangle=\frac{L_{a_4}}{||L_{a_4}||}=\frac{1}{\sqrt{4|a|^2+3}}[a(|0000\rangle+|0101\rangle+|1010\rangle+|1111\rangle)+\mathrm{i}|0001\rangle+|0110\rangle-\mathrm{i}|1011\rangle],$$
then
\begin{equation}\small
C_{2-\mathrm{ME}}(|\psi_{5}\rangle)=\min\{\frac{2\sqrt{4|a|^4+6|a|^2+2}}{4|a|^2+3},\frac{2\sqrt{12|a|^2+2}}{4|a|^2+3}\}.
\end{equation}
If~$|a|^2\leq\frac{3}{2}$, then
\begin{equation}
C_{2-\mathrm{ME}}(|\psi_{5}\rangle)=\frac{2\sqrt{4|a|^4+6|a|^2+2}}{4|a|^2+3};
\end{equation}
if $|a|^2>\frac{3}{2}$,
 \begin{equation}
C_{2-\mathrm{ME}}(|\psi_{5}\rangle)=\frac{2\sqrt{12|a|^2+2}}{4|a|^2+3}.
\end{equation}

The 3-ME concurrence of $|\psi_{5}\rangle$ is
\begin{equation}\small
C_{3-\mathrm{ME}}(|\psi_{5}\rangle)=\min\{\sqrt{\frac{7}{6}-\frac{7+24|a|^2}{6(4|a|^2+3)^2}},~\sqrt{\frac{4}{3}-\frac{2(16|a|^4+6)}{3(4|a|^2+3)^2}}~\}.\end{equation}
If $\frac{3-\sqrt{3}}{6}~\leq|a|^2~\leq\frac{3+\sqrt{3}}{6}$,
 \begin{equation}C_{3-\mathrm{ME}}(|\psi_{5}\rangle)=\sqrt{\frac{7}{6}-\frac{7+24|a|^2}{6(4|a|^2+3)^2}};\end{equation}
if~$|a|^2>\frac{3+\sqrt{3}}{6}$~or~$0\leq|a|^2\leq\frac{3-\sqrt{3}}{6}$,\begin{equation}C_{3-\mathrm{ME}}(|\psi_{5}\rangle)=\sqrt{\frac{4}{3}-\frac{2(16|a|^4+6)}{3(4|a|^2+3)^2}}.\end{equation}
The 4-ME concurrence of $|\psi_{5}\rangle$ is
\begin{equation}
C_{4-\mathrm{ME}}(|\psi_{5}\rangle)=\sqrt{1-\frac{1}{(4|a|^2+3)^2}}.
\end{equation}
The negativities for $\rho=|\psi_{5}\rangle\langle\psi_{5}|$ are
\begin{equation}
N^{1}=N^{2}=N^{3}=N^{4}=\sqrt{1-\frac{1}{(4|a|^2+3)^2}}.
\end{equation}
It implies that
if~$|a|^2\leq\frac{3}{2}$, then \begin{equation}C_{2-\mathrm{ME}}(|\psi_{5}\rangle)=\min\{N^{1},N^{2},N^{3},N^{4}\}.\end{equation}

(\uppercase\expandafter{\romannumeral6})~~Let
\begin{equation*}|\psi_{4}\rangle=\frac{L_{ab_3}}{\|L_{ab_3}\|}=\frac{1}{\sqrt{3|a|^2+|b|^2+2}}
[a(|0000\rangle+|1111\rangle)+\frac{a+b}{2}(|0101\rangle+|1010\rangle)].\end{equation*}
By some tedious computing, one can get the $k$-ME concurrence ($k=2,3$) for $|\psi_4\rangle$ are
 \begin{equation}\small\begin{array}{rl}\label{45}
  C_{2-\mathrm{ME}}(|\psi_{4}\rangle)=&\min\{\sqrt{1-\frac{1+8|a|^2}{(3|a|^2+|b|^2+2)^2}}, \sqrt{2-\frac{M+1+8|a|^2}{(3|a|^2+|b|^2+2)^2}}\},\\
C_{3-\mathrm{ME}}(|\psi_{4}\rangle)=&\sqrt{\frac{4}{3}-\frac{M+3+24|a|^2}{3(3|a|^2+|b|^2+2)^2}},\\
C_{4-\mathrm{ME}}(|\psi_{4}\rangle)=&\sqrt{1-\frac{1+8|a|^2}{(3|a|^2+|b|^2+2)^2}},
\end{array}\end{equation}
 where \begin{equation*}\small\begin{array}{rl}
M=&\max\{~6|a|^4+2|b|^4+8|a|^2+3,\\
&~~~~~~~~~\frac{|a+b|^4}{4}+\frac{|a-b|^4}{4}+|3a+b|^2+|a-b|^2+4|a|^4-2|a|^2+2|b|^2+2\\
&~~~~~~~~~+2|a|^2|a+b|^2+[(a+b)a^{\ast}+a(a+b)^{\ast}+1]^2,\\
&~~~~~~~~~\frac{|a+b|^4}{4}+\frac{|a-b|^4}{4}+|3a-b|^2+|a+b|^2+4|a|^4-2|a|^2+2|b|^2+2\\
&~~~~~~~~~+2|a|^2|a-b|^2+[(a-b)a^{\ast}+a(a-b)^{\ast}+1]^2\}.
\end{array}
\end{equation*}
The negativities of $|\psi_4\rangle$ are
\begin{equation}\small
\label{46}
N^{1}=N^{2}=N^{3}=N^{4}=\sqrt{1-\frac{1+8|a|^2}{(3|a|^2+|b|^2+2)^2}}.
\end{equation}Hence if~$(3|a|^2+|b|^2+2)^2\geq M$,
\begin{equation*}
C_{2-\mathrm{ME}}(|\psi_{4}\rangle)=\min\{N^{1},N^{2},N^{3},N^{4}\}.
\end{equation*}

(\uppercase\expandafter{\romannumeral 7})~~For $L_{a_2b_2}$, let
\begin{equation*}\small
|\psi_{3}\rangle=\frac{L_{a_2b_2}}{||L_{a_2b_2}||}=\frac{1}{\sqrt{2|a|^2+2|b|^2+2}}[a(|0000\rangle+|1111\rangle)+b(|0101\rangle+|1010\rangle)+|0110\rangle+|0011\rangle].
\end{equation*}
Simple algebra shows that
\begin{equation}\small
C_{2-\mathrm{ME}}(|\psi_{3}\rangle)=\min\{\sqrt{1-\frac{1}{(|a|^2+|b|^2+1)^2}},\sqrt{1-\frac{1}{(|a|^2+|b|^2+1)^2}+\frac{2|a|^2+2|b|^2-(ab^{\ast}+a^{\ast}b)^2}{(|a|^2+|b|^2+1)^2}}\},
\end{equation}
\begin{equation}
\begin{array}{rl}
C_{3-\mathrm{ME}}(|\psi_{3}\rangle)=&\min\{\sqrt{1-\frac{1}{3(|a|^2+|b|^2+1)^2}+\frac{2|a|^2|b|^2}{3(|a|^2+|b|^2+1)^2}},\\
&~~~~~~~~~\sqrt{1-\frac{1}{(|a|^2+|b|^2+1)^2}+\frac{2|a|^2+2|b|^2-(a^{\ast}b+ab^{\ast})^2}{3(|a|^2+|b|^2+1)^2}}~\},~~~~~
\end{array}
\end{equation}
\begin{equation}
C_{4-\mathrm{ME}}(|\psi_{3}\rangle)=\sqrt{1-\frac{1}{2(|a|^2+|b|^2+1)^2}}.
\end{equation}
The negativities of $|\psi_{3}\rangle$ are
\begin{equation}
\begin{array}{rl}
N^{1}=&N^{3}=\sqrt{1-\frac{1}{(|a|^2+|b|^2+1)^2}},\\
N^{2}=&N^{4}=1,
\end{array}
\end{equation}
Therefore, for $|\psi_{3}\rangle$ one has
\begin{equation*}
C_{2-\mathrm{ME}}(|\psi_{3}\rangle)=\min\{N^{1},N^{2},N^{3},N^{4}\}
\end{equation*}
if~$2|a|^2+2|b|^2-(ab^{\ast}+a^{\ast}b)^2\geq0$.\\

(\uppercase\expandafter{\romannumeral 8})~~For state $L_{abc_2}$, let
\begin{equation*}
\begin{array}{rl}
|\psi_{2}\rangle=&\frac{L_{abc_2}}{||L_{abc_2}||}\\
=&\frac{1}{\sqrt{|a|^2+|b|^2+2|c|^2+1}}[\frac{a+b}{2}(|0000\rangle+|1111\rangle)+\frac{a-b}{2}(|0011\rangle+|1100\rangle)\\
&+c(|0101\rangle+|1010\rangle)+|0110\rangle].
\end{array}
\end{equation*}
Then\begin{equation}\begin{array}{rl}
C_{2-\mathrm{ME}}(|\psi_{2}\rangle)=\min\{&\sqrt{1-\frac{1}{(|a|^2+|b|^2+2|c|^2+1)^2}}, ~\sqrt{\frac{4(|c|^4+|a|^2|b|^2+2|a|^2|c|^2+2|b|^2|c|^2+|a|^2+|b|^2)}{(|a|^2+|b|^2+2|c|^2+1)^2}},\\
&\sqrt{1-\frac{2-t+(ab^{\ast}+a^{\ast}b)^2+(ab^{\ast}+a^{\ast}b)(8|c|^2-4)+2(a+b)^2(c^{\ast})^2+2[(a+b)^{\ast}]^2c^2}{2(|a|^2+|b|^2+2|c|^2+1)^2}},\\
&\sqrt{1-\frac{2-t+(ab^{\ast}+a^{\ast}b)^2-(ab^{\ast}+a^{\ast}b)(8|c|^2-4)+2(a-b)^2(c^{\ast})^2+2[(a-b)^{\ast}]^2c^2}{2(|a|^2+|b|^2+2|c|^2+1)^2}}~\},
\end{array}\end{equation}
\begin{equation}\begin{array}{rl}
C_{3-\mathrm{ME}}(|\psi_{2}\rangle)=\min\{&\sqrt{\frac{2}{3}+\frac{4(|c|^4+|a|^2|b|^2+2|a|^2|c|^2+2|b|^2|c|^2+|a|^2+|b|^2)-2}{3(|a|^2+|b|^2+2|c|^2+1)^2}},\\
&\sqrt{1-\frac{6-t+(ab^{\ast}+a^{\ast}b)^2+(ab^{\ast}+a^{\ast}b)(8|c|^2-4)+2(a+b)^2(c^{\ast})^2+2[(a+b)^{\ast}]^2c^2}{6(|a|^2+|b|^2+2|c|^2+1)^2}},\\
&\sqrt{1-\frac{6-t+(ab^{\ast}+a^{\ast}b)^2-(ab^{\ast}+a^{\ast}b)(8|c|^2-4)+2(a-b)^2(c^{\ast})^2+2[(a-b)^{\ast}]^2c^2}{6(|a|^2+|b|^2+2|c|^2+1)^2}}~\},
\end{array}\end{equation}
\begin{equation}\small
C_{4-\mathrm{ME}}(|\psi_{2}\rangle)=\sqrt{1-\frac{1}{(|a|^2+|b|^2+2|c|^2+1)^2}},
\end{equation}
where $t=|a|^4+|b|^4+2|a|^2|b|^2+8|c|^2$.\\

The negativities are
\begin{equation}\small
N^{1}=N^{2}=N^{3}=N^{4}=\sqrt{1-\frac{1}{(|a|^2+|b|^2+2|c|^2+1)^2}}.
\end{equation}
Thus one can get relation for $|\psi_{2}\rangle$
\begin{equation*}
C_{2-\mathrm{ME}}(|\psi_{2}\rangle)=\min\{N^{1},N^{2},N^{3},N^{4}\}
\end{equation*}
if the coefficients of  $|\psi_{2}\rangle$ satisfy following inequation
\begin{equation}
\begin{array}{rl}\begin{cases}\label{51}
t-(ab^{\ast}+a^{\ast}b)^2&\geq(8|c|^2-4)(ab^{\ast}+ba^{\ast})+2(a+b)^2(c^{\ast})^2+2[(a+b)^{\ast}]^2c^2\\
t-(ab^{\ast}+a^{\ast}b)^2&\geq(-8|c|^2+4)(ab^{\ast}+ba^{\ast})+2(a-b)^2(c^{\ast})^2+2[(a-b)^{\ast}]^2c^2\\
|a|^4+|b|^4+4|c|^2&\leq2|a|^2|b|^2+4|a|^2|c|^2+4|c|^2|b|^2+2|a|^2+2|b|^2.\\
 \end{cases}\end{array}
\end{equation}

(\uppercase\expandafter{\romannumeral 9})~~Given $G_{abcd}$, let
\begin{equation*}
\begin{array}{rl}
|\psi_{1}\rangle=&\frac{G_{abcd}}{\|G_{abcd}\|}\\=&\frac{1}{\sqrt{|a^2|+|b|^2+|c|^2+|d|^2}}[\frac{a+d}{2}(|0000\rangle+|1111\rangle)+\frac{a-d}{2}(|0011\rangle+|1100\rangle)\\
&+\frac{b+c}{2}(|0101\rangle+|1010\rangle)+\frac{b-c}{2}(|0110\rangle+|1001\rangle)].\\
\end{array}
\end{equation*}
By tedious computing, we derive
\begin{equation}\begin{array}{rl}\label{56}
C_{2-\mathrm{ME}}(|\psi_{1}\rangle)=\min\{&\sqrt{2-\frac{2(|a|^4+|b|^4+|c|^4+|d|^4)}{(|a|^2+|b|^2+|c|^2+|d|^2)^2}}, \sqrt{2-\frac{x}{4(|a|^2+|b|^2+|c|^2+|d|^2)^2}},\\
&\sqrt{2-\frac{y}{4(|a|^2+|b|^2+|c|^2+|d|^2)^2}}, ~1~\},
\end{array}\end{equation}
\begin{equation}\begin{array}{rl}
C_{3-\mathrm{ME}}(|\psi_{1}\rangle)=\min\{&\sqrt{\frac{4}{3}-\frac{2(|a|^4+|b|^4+|c|^4+|d|^4)}{3(|a|^2+|b|^2+2|c|^2+1)^2}}, \sqrt{\frac{4}{3}-\frac{x}{12(|a|^2+|b|^2+|c|^2+|d|^2)^2}},\\
&\sqrt{\frac{4}{3}-\frac{y}{12(|a|^2+|b|^2+|c|^2+|d|^2)^2}}~\},
\end{array}\end{equation}
\begin{equation}
\label{57}
C_{4-\mathrm{ME}}(|\psi_{1}\rangle)=1,
\end{equation}
\begin{equation}\label{58}
N^{1}=N^{2}=N^{3}=N^{4}=1.
\end{equation}
Here
\begin{equation*}
\begin{array}{rl}
x\equiv&(|a+d|^2+|b+c|^2)^2+(|a-d|^2+|b-c|^2)^2+[(a+d)(b+c)^{\ast}+(b+c)(a+d)^{\ast}]^2\\
&+[(a-d)(b-c)^{\ast}+(b-c)(a-d)^{\ast}]^2,\\
y\equiv&(|a+d|^2+|b-c|^2)^2+(|a-d|^2+|b+c|^2)^2+[(a+d)(b-c)^{\ast}+(b-c)(a+d)^{\ast}]^2\\
&+[(a-d)(b+c)^{\ast}+(b+c)(a-d)^{\ast}]^2.
\end{array}
\end{equation*}
Using  Eqs.$\eqref{56}$ and $\eqref{58}$, one can easily get
\begin{equation*}
C_{2-\mathrm{ME}}(|\psi_{1}\rangle)=\min\{N^{1},N^{2},N^{3},N^{4}\}
\end{equation*}
in case of  the following inequality holds
\begin{equation}
\label{55}
(|a|^2+|b|^2+|c|^2+|d|^2)^2\geq M,
\end{equation}
 where $M=\max\{2(|a|^4+|b|^4+|c|^4+|d|^4),~\frac{1}{4}x,~\frac{1}{4}y\}$.\\

From above calculating we know the relation between 2-ME concurrence and negativity
\begin{equation}\label{9}
C_{2-\mathrm{ME}}(|\psi_{i}\rangle)=\min\{N^{1},N^{2},N^{3},N^{4}\}
\end{equation}
holds for four families of four-qubit states, while for the rest five families of four-qubit states Eq.$\eqref{9}$ is true under certain conditions. That is, Eqs.$\eqref{9}$ doesn't hold for all pure 4-qubit states unless the coefficients satisfy certain conditions.
\section{Relationship between $k$-ME Concurrence and Tangle for $n$-qubit~W~state}
 W-class states are important physical resource and have attracted much attention. They have been used for many vital quantum information processing tasks \cite{PLA1,NJP1,55, PRA65.0323108}, such as a quantum channel for teleportation \cite{PLA1,NJP1}, quantum key distribution \cite{55}.
 In this section,  we focus on the relation between two entanglement measures, $k$-ME concurrence and tangle $\tau_{ij}$ for W state.\\

For an $n$-qubit~W-class~state
\begin{equation}|\psi\rangle=a_1|00\cdots01\rangle+a_2|00\cdots10\rangle+\cdots+a_{n-1}|01\cdots00\rangle+a_n|10\cdots00\rangle,\end{equation} where~$a_i$~are complex numbers with~$\sum\limits_i|a_i|^2=1$, then one have
\begin{equation}
\begin{array}{rl}
1-\mathrm{Tr}(\rho_{n}^2)=&2|a_1|^2|a_2|^2+2|a_1|^2|a_3|^2+\cdots+2|a_1|^2|a_n|^2=~~2\sum\limits_{i=2}^{n}|a_1|^2|a_i|^2,
\end{array}
\end{equation}
\begin{equation}
\begin{array}{rl}
1-\mathrm{Tr}(\rho_{(n-1)n}^2)=&2|a_1|^2|a_{3}|^2+2|a_2|^2|a_{3}|^2+\cdots+2|a_{1}|^2|a_{n}|^2+2|a_{2}|^2|a_{n}|^2\\
=&2\sum\limits_{i=3}^{n}|a_i|^2(|a_1|^2+|a_2|^2),
\end{array}
\end{equation}
$$\cdots~~~~~~~~~~~~~~~~~~~~~~~~~~~~~~~~\cdots~~~~~~~~~~~~~~~~~~~~~~~~~~~~~~~~~~~~~~~~~~~~~~~$$
\begin{equation}
\begin{array}{rl}
1-\mathrm{Tr}(\rho_{12\cdots[\frac{n}{2}]}^2)=&2|a_{1}|^2|a_{[\frac{n+1}{2}]+1}|^2+\cdots+2|a_{1}|^2|a_{n-1}|^2+2|a_{1}|^2|a_{n}|^2+\cdots\\
&+2|a_{[\frac{n+1}{2}]}|^2|a_{[\frac{n+1}{2}]+1}|^2+\cdots+2|a_{[\frac{n+1}{2}]}|^2|a_{n-1}|^2+2|a_{[\frac{n+1}{2}]}|^2|a_{n}|^2~~~~~~~~~~~~~~~\\
=&2\sum_{i=[\frac{n+1}{2}]+1}^{n}|a_i|^2(|a_1|^2+|a_2|^2+\cdots+|a_{[\frac{n+1}{2}]}|^2).~~~~~~~~~~~~~~~~~~~~~~~~~~~~~
\end{array}
\end{equation}
Here  $\rho=|\psi\rangle\langle\psi|$ is the $n$-qubit W-class state, and $\rho_A=\mathrm{Tr}_{\bar{A}}\rho$ is the reduced density matrix of subsystem $A$ ($\bar{A}$ is the complement of $A$ in $\{1,2,\cdots,n\}$).
We  obtain the 2-tangle $\tau_{ij}$~\cite{37}~of W class state $\rho=|\psi\rangle\langle\psi|$
 \begin{equation}\tau_{ij}=4|a_{n+1-i}|^2|a_{n+1-j}|^2.
\end{equation}

Specially given~the $n$-qubit W state $$|W\rangle=\frac{1}{\sqrt{n}}(|00\cdots01\rangle+|00\cdots10\rangle+\cdots+|01\cdots00\rangle+|10\cdots00\rangle),$$
using inequality
\begin{equation}
t(n-t)\geq(t-1)[n-(t-1)]~,~~~t\leq[\frac{n+1}{2}],~n\geq3,
\end{equation} one can prove that for W state $|W\rangle$, the following equality holds,
\begin{equation*}
C_{k-\mathrm{ME}}(|W\rangle)=\sqrt{\frac{2}{k}\{[(k-1)n-\frac{k(k-1)}{2}]\tau_{ij}\}},~~~k=2,~3,~\cdots,~n,
\end{equation*}
which shows the relation between $k$-ME concurrence and 2-tangle of $n$-qubit W-state.
\section{Conclusion}
In summary, we show that for an $n$-qubit state $\rho$, the $k$-ME ($k$=$n$) concurrence is not less than the quadratic mean of negativities $N^{i}$ ($i=1, 2, \cdots, n$) with equality for all $n$-qubit pure states, demonstrate that our bound about $n$-ME concurrence is exact for some mixed states,   and reveal the relations between $k$-ME concurrence and negativity, between $k$-ME concurrence and  polynomial invariants for $3$-qubit and $4$-qubit pure states. We prove that the $2$-ME concurrence of any $3$-qubit state $|\psi\rangle$  equals to the minimum of the negativities ($N^{i}$, $i=1, 2, 3$) of $|\psi\rangle$. Meanwhile the $2$-ME concurrence and negativity of the nine families of $4$-qubit pure states have  similar relations. Besides, we also exhibit strong links between $k$-ME concurrence and polynomial invariants for three and four qubit states, and between $k$-ME concurrence and tangle for $n$-qubit W-state. From these clear quantitative connections between entanglement measures and polynomial invariants, one may get more information of entanglement ways of  states.
\begin{acknowledgments}
This work was supported by the National Natural Science Foundation of China under Grant No. 11475054 and the Hebei Natural Science Foundation  under Grant Nos. A2016205145 and A2018205125.
\end{acknowledgments}


\begin{thebibliography}{99}


\bibitem{RMP81.865} R. Horodecki, \textit{et al.}, Quantum entanglement, \href{http://journals.aps.org/rmp/abstract/10.1103/RevModPhys.81.865} {Rev. Mod. Phys. \textbf{81}, 865 (2009).}

\bibitem{PRL67.661} A. K. Ekert, Quantum cryptography based on Bell's theorem, \href{http://journals.aps.org/prl/abstract/10.1103/PhysRevLett.67.661} {Phys. Rev. Lett. \textbf{67}, 661 (1991).}

\bibitem{PRL70.1895} C. H. Bennett, \textit{et al.}, Teleporting an unknown quantum state via dual classical and Einstein-Podolsky-Rosen channels, \href{http://journals.aps.org/prl/abstract/10.1103/PhysRevLett.70.1895} {Phys. Rev. Lett. \textbf{70}, 1895 (1993).}

\bibitem{EPL84.50001} T. Gao, F. L. Yan, and Y. C. Li, Optimal controlled teleportation, \href{http://iopscience.iop.org/article/10.1209/0295-5075/84/50001/meta} {Europhys. Lett. \textbf{84}, 50001 (2008).}

\bibitem{PRL69.2881} C. H. Bennett and S. J. Wiesner,  Communication via one- and two-particle operators on Einstein-Podolsky-Rosen states, \href{http://journals.aps.org/prl/abstract/10.1103/PhysRevLett.69.2881} {Phys. Rev. Lett. \textbf{69}, 2881 (1992).}

\bibitem{Nature2010} C. Gross, \textit{et al.}, Nonlinear atom interferometer surpasses classical precision limit, \href{https://doi.org/10.1038/nature08919} {Nature. \textbf{464}, 1165 (2010).}

\bibitem{JPA.47.424005} C. Eltschka and J. Siewert, Quantifying entanglement resources, \href{http://iopscience.iop.org/article/10.1088/1751-8113/47/42/424005/meta} {J. Phys. A: Math. Theor. \textbf{47}, 424005 (2014).}

\bibitem{7}  C. H. Bennett, \textit{et al.}, Mixed-state entanglement and quantum error correction,  \href{https://journals.aps.org/pra/abstract/10.1103/PhysRevA.54.3824} {Phys. Rev. A \textbf{54}, 3824 (1996).}

\bibitem{8}  S. Hill and W. K. Wootters, Entanglement of a pair of quantum bits, \href{http://journals.aps.org/prl/abstract/10.1103/PhysRevLett.78.5022} { Phys. Rev. Lett. \textbf{78}, 5022 (1997).}

\bibitem{9}W. K. Wootters, Entanglement of formation of an arbitrary state of two qubits,   \href{http://journals.aps.org/prl/abstract/10.1103/PhysRevLett.80.2245} {Phys. Rev. Lett. \textbf{80}, 2245 (1998).}

\bibitem{10}P. Rungta, \textit{et al.}, Universal state inversion and concurrence in arbitrary dimensions, \href{http://journals.aps.org/pra/abstract/10.1103/PhysRevA.64.042315}  {Phys. Rev. A \textbf{64}, 042315 (2001).}

\bibitem{11}A. R. R. Carvalho, F. Mintert, and A. Buchleitner, Decoherence and multipartite entanglement, \href{http://journals.aps.org/prl/abstract/10.1103/PhysRevLett.93.230501} { Phys. Rev. Lett. \textbf{93}, 230501 (2004).}

\bibitem{18} Y. Hong, T. Gao, and F. L. Yan, Measure of multipartite entanglement with computable lower bounds, \href{https://journals.aps.org/pra/abstract/10.1103/PhysRevA.86.062323}   {Phys. Rev. A \textbf{86}, 062323 (2012).}

\bibitem{17}  T. Gao and Y. Hong, Detection of genuinely entangled and nonseparable $n$-partite quantum states, \href{https://journals.aps.org/pra/abstract/10.1103/PhysRevA.82.062113}  {Phys. Rev. A \textbf{82}, 062113 (2010).}

\bibitem{40} T. Gao, \textit{et al.}, Efficient $k$-separability criteria for mixed multipartite quantum states, \href{http://iopscience.iop.org/article/10.1209/0295-5075/104/20007/meta}  { Europhys. Lett. \textbf{104}, 20007 (2013).}

\bibitem{41} T. Gao, F. L. Yan, and S. J. van Enk, Permutationally invariant part of a density matrix and nonseparability of $N$-qubit states, \href{http://journals.aps.org/prl/abstract/10.1103/PhysRevLett.112.180501}  {Phys. Rev. Lett. \textbf{112}, 180501 (2014).}

\bibitem{PRA65.032314}G. Vidal and R. F. Werner, Computable measure of entanglement, \href{https://journals.aps.org/pra/abstract/10.1103/PhysRevA.65.032314} {Phys. Rev. A \textbf{65}, 032314 (2002).}

\bibitem{PRA61.062313} W. D\"{u}r, \textit{et al.}, Distillability and partial transposition in bipartite systems, \href{https://journals.aps.org/pra/abstract/10.1103/PhysRevA.61.062313} {Phys. Rev. A \textbf{61}, 062313 (2000).}

\bibitem{PRL80.5239} M. Horodecki, P. Horodecki, and R. Horodecki, Mixed-state entanglement and distillation: is there a ``bound" entanglement in nature? \href{http://journals.aps.org/prl/abstract/10.1103/PhysRevLett.80.5239} {Phys. Rev. Lett. \textbf{80}, 5239 (1998).}

\bibitem{PRA68.062304} S. Lee, \textit{et al.}, Convex-roof extended negativity as an entanglement measure for bipartite quantum systems, \href{https://journals.aps.org/pra/abstract/10.1103/PhysRevA.68.062304} {Phys. Rev. A \textbf{68}, 062304 (2003).}
\bibitem{JM047}G. Vidal, Entanglement monotones, \href{http://www.tandfonline.com/doi/abs/10.1080/09500340008244048} { J. Mod. Opt. \textbf{47}, 355 (2000)}.

\bibitem{PRA77.042117}S. S. Sharma and N. K. Sharma, Quantum coherences, $K$-way negativities and multipartite entanglement, \href{https://journals.aps.org/pra/abstract/10.1103/PhysRevA.77.042117} {Phys. Rev. A \textbf{77}, 042117 (2008).}

\bibitem{pra62062314}  W. D\"{u}r, G. Vidal, and J. I. Cirac, Three qubits can be entangled in two inequivalent ways, \href{https://journals.aps.org/pra/abstract/10.1103/PhysRevA.62.062314} {Phys. Rev. A \textbf{62}, 062314 (2000).}

\bibitem{pra82032121} B. Kraus, Local unitary equivalence and entanglement of multipartite pure states, \href{https://journals.aps.org/pra/abstract/10.1103/PhysRevA.82.032121} {Phys. Rev. A \textbf{82}, 032121 (2010).}

\bibitem{prl108050501}B. Liu, \textit{et al.}, Local unitary classification of arbitrary dimensional multipartite pure states, \href{http://journals.aps.org/prl/abstract/10.1103/PhysRevLett.108.050501} {Phys. Rev. Lett. \textbf{108}, 050501 (2012).}

\bibitem{PRA.87}S. H. Wang, Y. Lu, and G. L. Long, Entanglement classification of $2\times 2 \times 2 \times d$ quantum systems via the ranks of the multiple coefficient matrices, \href{http://journals.aps.org/pra/abstract/10.1103/PhysRevA.87.062305} {Phys. Rev. A \textbf{87}, 062305 (2013).}

\bibitem{PRA.88} X. R. Li and D. F. Li, Polynomial invariants of degree $4$ for even-$n$ qubits and their applications
in entanglement classification, \href{http://journals.aps.org/pra/abstract/10.1103/PhysRevA.88.022306} {Phys. Rev. A \textbf{88}, 022306 (2013).}

\bibitem{JPA.50.2017}M. Sanz, \textit{et al.}, Entanglement classification with algebraic geometry, \href{https://doi.org/10.1088/1751-8121/aa6926} {J. Phys. A: Math. Theor. \textbf{50}, 195303 (2017). }

\bibitem{PRA.58.1833}M. Grassl, M. R\"{o}tteler, and T. Beth, Computing local invariants of quantum-bit  systems, \href{http://journals.aps.org/pra/abstract/10.1103/PhysRevA.58.1833} {Phys. Rev. A \textbf{58}, 1833 (1998).}
\bibitem{20} N. Linden \textit{et al.}, On multi-particle entanglement, \href{https://doi.org/10.1002/(SICI)1521-3978(199806)46:4/5<567::AID-PROP567>3.0.CO;2-H} {Fortsch. Phys. \textbf{46}, 567 (1998).}
\bibitem{JMP43}D. A. Meyer, and N. R. Wallach, Global entanglement in multiparticle systems, \href{https://doi.org/10.1063/1.1497700} {J. Math. Phys. \textbf{43}, 4273 (2002).}
\bibitem{pra92042307}J. H. Choi, and J. S. Kim, Negativity and strong monogamy of multiparty quantum entanglement beyond qubits,  \href{https://journals.aps.org/pra/abstract/10.1103/PhysRevA.92.042307} {Phys. Rev. A \textbf{92}, 042307 (2015).}
\bibitem{37}  V. Coffman, J. Kundu, and W. K. Wootters, Distributed entanglement, \href{https://journals.aps.org/pra/abstract/10.1103/PhysRevA.61.052306} {Phys. Rev. A \textbf{61}, 052306 (2000).}

\bibitem{12} M. A. Nielsen and I. L. Chuang, \textit{Quantum Computation and Quantum Information} (Cambridge University Press, Cambridge, 2010).
\bibitem{PRA61042314}W. D\"{u}r, and J. I. Cirac,  Classification of multiqubit mixed states: separability and distillability properties, \href{https://journals.aps.org/pra/abstract/10.1103/PhysRevA.61.042314} {Phys. Rev. A \textbf{61}, 042314 (2000).}
\bibitem{GaoEPJD2011}  T. Gao and Y. Hong, Separability criteria for several classes of $n$-partite
quantum states,   {Eur. Phys. J. D \textbf{61}, 765 (2011).}
\bibitem{36}  A. Sudbery, On local invariants of pure three-qubit states, \href{http://iopscience.iop.org/article/10.1088/0305-4470/34/3/323/meta} {J. Phys. A: Math. Gen. \textbf{34}, 643 (2001).}

\bibitem{jpa346787} H. Barnum and N. Linden, Monotones and invariants for multi-particle quantum states, \href{http://iopscience.iop.org/article/10.1088/0305-4470/34/35/305/meta} {J. Phys. A: Math. Gen. \textbf{34}, 6787 (2001).}


\bibitem{PRA65.2002.052112}  F. Verstraete, \textit{et al.}, Four qubits can be entangled in nine different ways, \href{https://journals.aps.org/pra/abstract/10.1103/PhysRevA.65.052112} {Phys. Rev. A \textbf{65}, 052112 (2002).}

\bibitem{PLA1}V. N. Gorbachev, \textit{et al.}, Can the states of the W-class be suitable for teleportation? \href{https://doi.org/10.1016/S0375-9601(03)00906-X} {Phys. Lett. A \textbf{314}, 267 (2003).}

\bibitem{NJP1}J. Joo, \textit{et al.}, Quantum teleportation via a W state, \href{http://iopscience.iop.org/article/10.1088/1367-2630/5/1/136/meta} {New J. Phys. \textbf{5}, 136 (2003).}

\bibitem{55}J. Joo, \textit{et al.},  Quantum secure communication with W states, \href{https://arxiv.org/abs/quant-ph/0204003} {arXiv:quant-ph/0204003}.

\bibitem{PRA65.0323108}A. Cabello, Bell's theorem with and without inequalities for the three-qubit Greenberger-Horne-Zeilinger and W states, \href{https://journals.aps.org/pra/abstract/10.1103/PhysRevA.65.032108} {Phys. Rev. A \textbf{65}, 032108 (2002).}

\end{thebibliography}
\end{document}